\documentclass[10pt]{article}
\usepackage{makeidx}

\makeindex
\newtheorem{definition}{Definition}

\begin{document}

\def\today{27~November~2002}
\def \Oeuvres{O$\!$euvres}
\def \Lecons{Le\c cons}
\def \ie {i.e.~}
\def \LHS{l.h.s.~}
\def \formerDT{Darboux involution}
\def\CRAS{C.~R.~Acad.~Sc.~Paris}
\def\SAM{Stud.~Appl.~Math.~}
\def\AnnENS{Ann.~\'Ec.~Norm.~}
\def\Pn     {{\rm Pn}}
\def\PI     {{\rm P1}}
\def\PII    {{\rm P2}}
\def\PIII   {{\rm P3}}
\def\PIV    {{\rm P4}}
\def\PV     {{\rm P5}}
\def\PVI    {{\rm P6}}

\def\pasq{q} 
\def\pas{h}  

\def\Alpha{A}
\def\Beta {B}
\def\abcd{\alpha,\beta,\gamma,\delta}

\def \ccomma{\raise 2pt\hbox{,}} 
\def \D {\hbox{d}}
\def \Log {\mathop{\rm Log}\nolimits}
\def \sinh{\mathop{\rm sinh}\nolimits}
\def \sech{\mathop{\rm sech}\nolimits}
\def \grad{\mathop{\rm grad}\nolimits}
\def \mod#1{\vert #1 \vert}
\def \bfE {{    E}}
\def \bfR {{    R}}
\def \bfp {{    p}}
\def \bfu {{    u}}

\title{
\textbf{The Painlev\'e methods}
\footnote{ \textit{Nonlinear integrable systems: classical and quantum},
ed.~A.~Kundu, 
Special issue, Proceedings of Indian Science Academy, A, to appear. 
S2002/066. nlin.SI/0211048} }

\author{
\\
\\
\\
{\bf R.~Conte}\dag \  and {\bf M.~Musette}\ddag
\\
\\
\\
\noindent
\dag Service de physique de l'\'etat condens\'e,
CEA--Saclay,
\\
\noindent
F-91191 Gif-sur-Yvette Cedex
\\
\\
\noindent
\ddag Dienst Theoretische Natuurkunde,
Vrije Universiteit Brussel,
Pleinlaan 2,
\\
\noindent
B-1050 Bruxelles
\\
\\
\\
}

\maketitle

\tableofcontents
\vfill \eject

This short review is an introduction to a great variety of methods,
the collection of which is called the
\textit{Painlev\'e analysis},
 \index{Painlev\'e!analysis}
intended at producing all kinds of exact (as opposed to perturbative)
results on nonlinear equations, whether ordinary, partial, or discrete.

\section{The classical program of the Painlev\'e school and its achievements}
\label{sectionTheclassicalprogram}

It is impossible to understand anything to the Painlev\'e property without
keeping in mind the original problem as stated by L.~Fuchs, Poincar\'e and
Painlev\'e:
to \textit{define new functions} from ordinary differential equations (ODEs).
This simply formulated problem implies to select those ODEs whose
general solution can be made singlevalued
by some uniformization procedure (cuts, Riemann surface),
so as to fit the definition of a \textit{function}.
This property (the possibility to uniformize the general solution of an ODE),
nowadays called the
\textit{Painlev\'e property} (PP),
 \index{Painlev\'e!property}
is equivalent to the more practical definition.

\begin{definition}
The \textit{Painlev\'e property} of an ODE is
the absence of movable critical singularities in its general solution.
\index{Painlev\'e!property for ODEs}
\end{definition}

Let us recall that a singularity is said \textit{movable}
(as opposed to \textit{fixed})
 \index{singularity!movable}
 \index{singularity!fixed}
if its location depends on the initial conditions,
and \textit{critical} if multivaluedness takes place around it.
Other definitions of the PP,
excluding for instance the essential singularities,
or replacing
``movable critical singularities''
by ``movable singularities other than poles'',
or ``its general solution'' by ``all its solutions'',
are incorrect.
Two examples taken from Chazy \cite{ChazyThese} explain why this is so.
The first example is the celebrated Chazy's equation of class III
\begin{eqnarray}
& &
u''' - 2 u u'' + 3 u'^2=0,
\label{eqChazyIII}
\end{eqnarray}
whose general solution is only defined inside or outside a circle
characterized by the three initial conditions (two for the center, one for the
radius);
this solution is holomorphic in its domain of definition and cannot be
analytically continued beyond it.
This equation therefore has the PP,
and the only singularity is a movable analytic essential singular line
which is a natural boundary.

The second example \cite[p.~360]{ChazyThese} is the third order second degree
ODE
\begin{eqnarray}
& &
(u''' -2 u' u'')^2 + 4 u''^2 (u'' - u'^2 - 1)=0,\
\end{eqnarray}
whose general solution is singlevalued,
\begin{eqnarray}
& &
u=e^{c_1 x + c_2}/c_1 + \frac{c_1^2-4}{4 c_1} x + c_3,
\end{eqnarray}
but which also admits a \textit{singular solution}
(envelope solution)
with a movable critical singularity,
\begin{eqnarray}
& &
u=C_2 - \Log \cos(x-C_1).
\end{eqnarray}
For more details,
see the arguments of Painlev\'e \cite[\S 2.6]{Cargese1996Conte}
and Chazy \cite[\S 5.1]{Cargese1996Conte}.

The PP is invariant under an arbitrary homography on the dependent variable
and an arbitrary change of the independent variable
(\textit{homographic group})
\begin{eqnarray}
& &
(u,x) \mapsto (U,X),\
u(x)=\frac{\alpha(x) U(X) + \beta(x)}{\gamma(x) U(X) + \delta(x)},\
X=\xi(x),\
\nonumber
\\
& &
 (\alpha, \beta, \gamma, \delta, \xi) \hbox{ functions},\
\alpha \delta - \beta \gamma \not=0.
\label{eqHomographicGroup}
\end{eqnarray}

Every linear ODE possesses the PP since its general solution depends linearly
on the movable constants,
so, in order to define new functions,
one must turn to nonlinear ODEs in a systematic way:
first order algebraic equations, then second order, \dots
The current achievements are the following.

First order algebraic ODEs (polynomial in $u,u'$, analytic in $x$)
define only one function, the
\textit{Weierstrass elliptic function} $\wp$,
new in the sense that its ODE
\begin{eqnarray}
& &
{u'}^2-4 u^3 + g_2 u + g_3=0,\
(g_2,g_3) \hbox{ arbitrary complex constants},
\label{eqWeierstrassOrder1}
\end{eqnarray}
is not reducible to a linear ODE.
Its only singularities are movable double poles.

Second order algebraic ODEs (polynomial in $u,u',u''$, analytic in $x$)
define six functions, the
\textit{Painlev\'e functions} $\Pn,n=1,\cdots,6$,
 \index{Painlev\'e!functions}
new because they are not reducible
to either a linear ODE or a first order ODE.
This question of \textit{irreducibility},
\index{irreducibility}
the subject of a long dispute between Painlev\'e and Joseph Liouville,
has been rigorously settled only recently \cite{U1990}.
The canonical representatives of $\PI$--$\PVI$ in their equivalence class
under the group (\ref{eqHomographicGroup}) are
\begin{eqnarray*}
\PI\ : \
u''
&=&
6 u^2 + x,
\\
\PII\ : \
u''
&=&
2 u^3 + x u + \alpha,
\\
\PIII\ 
: \
u''
&=&
\frac{u'^2}{u} - \frac{u'}{x} + \frac{\alpha u^2 + \gamma u^3}{4 x^2}
 + \frac{\beta}{4 x}
 + \frac{\delta}{4 u}\ccomma
\\
\PIV\ : \
u''
&=&
 \frac{u'^2}{2 u} +              \frac{3}{2} u^3 + 4 x u^2 + 2 x^2 u
- 2 \alpha u + \frac{\beta}{u}\ccomma
\\
\PV\ : \
u''
&=&
\left[\frac{1}{2 u} + \frac{1}{u-1} \right] {u'}^2
- \frac{u'}{x}
+ \frac{(u-1)^2}{x^2} \left[ \alpha u + \frac{\beta}{u} \right]
+ \gamma \frac{u}{x}
+ \delta \frac{u(u+1)}{u-1}\ccomma
\nonumber
\\
\PVI\ : \
u''
&=&
\frac{1}{2} \left[\frac{1}{u} + \frac{1}{u-1} + \frac{1}{u-x} \right] {u'}^2
- \left[\frac{1}{x} + \frac{1}{x-1} + \frac{1}{u-x} \right] u'
\\
& &
+ \frac{u (u-1) (u-x)}{x^2 (x-1)^2}
  \left[\alpha + \beta \frac{x}{u^2} + \gamma \frac{x-1}{(u-1)^2}
        + \delta \frac{x (x-1)}{(u-x)^2} \right]\ccomma
\end{eqnarray*}
in which $\abcd$ are arbitrary complex parameters.
Their only singularities are movable poles
(in the $e^x$ complex plane for $\PIII$ and $\PV$,
in the $x$ plane for the others),
with in addition three fixed critical singularities for $\PVI$,
located at $x=\infty,0,1$.

Third and higher order ODEs \cite{ChazyThese,BureauMII,Cos2000a,Cos2000b}
have not yet defined new functions.
Although there are some good candidates
(the Garnier system \cite{GarnierThese},
several fourth order ODEs \cite{KS1998,Cos2000a},
which all have a transcendental dependence on the constants of integration),
the question of their irreducibility
(to a linear, Weierstrass, or Painlev\'e equation)
is very difficult and still open.
To understand the difficulty, it is sufficient to consider the 
fourth order ODE for $u(x)$ defined by
\begin{eqnarray}
& &
u=u_1+u_2,\
u_1''=6 u_1^2+x,\
u_2''=6 u_2^2+x.
\end{eqnarray}
This ODE (easy to write by elimination of $u_1,u_2$)
has a general solution which depends transcendentally on the four constants 
of integration, and it is reducible.

The master equation $\PVI$ was first written 
by Picard in 1889 in a particular case, in a very elegant way.
Let $\varphi$ be the elliptic function defined by
\begin{eqnarray}
& &
\varphi\ :\ y \mapsto \varphi(y,x),\
y=\int_{\infty}^{\varphi} \frac{\D z}{\sqrt{z(z-1)(z-x)}},
\end{eqnarray}
and let $\omega_1(x),\omega_2(x)$ be its two half-periods.
Then the function
\begin{eqnarray}
& &
u\ :\ x \mapsto u(x)=\varphi(2 c_1 \omega_1(x) + 2 c_2 \omega_2(x),x),\
\end{eqnarray}
with $(c_1,c_2)$ arbitrary constants,
has no movable critical singularities,
and it satisfies a second order ODE which is
$\PVI$ in the particular case $\alpha=\beta=\gamma=\delta-1/2=0$.
The generic $\PVI$ was found simultaneously from two different approaches,
the nonlinear one of the Painlev\'e school as said above \cite{PaiCRAS1906},
and the linear one of R.~Fuchs \cite{FuchsP6} as an isomonodromy condition.
In the latter, one considers a second order linear ODE for $\psi(t)$
with four Fuchsian singularities of crossratio $x$
(located for instance at $t=\infty,0,1,x$),
with in addition, as prescribed by Poincar\'e for the isomonodromy problem,
one apparent singularity $t=u$,
\begin{eqnarray}
- \frac{2}{\psi} \frac{\D^2 \psi}{\D t^2}
& = &
  \frac{A}{t^2}
+ \frac{B}{(t-1)^2}
+ \frac{C}{(t-x)^2}
+ \frac{D}{t (t-1)}
+ \frac{3     }{4 (t-u)^2}
\nonumber
\\
& &
+ \frac{a     }{t (t-1) (t-x)}
+ \frac{b     }{t (t-1) (t-u)},
\end{eqnarray}
$(A,B,C,D)$ denoting constants and $(a,b)$ parameters.
The requirement that the monodromy matrix
(which transforms two independent solutions $\psi_1,\psi_2$
when $t$ goes around a singularity)
be independent of the nonapparent singularity $x$
results in the condition that $u$, as a function of $x$,
satisfies $\PVI$.

A useful by-product of this search for new functions is the
construction of several exhaustive lists (\textit{classifications})
 \index{classification}
of
second \cite{GambierThese,BureauMIII,CosODE2,CosScou,Cos1997},
third  \cite{ChazyThese,BureauMII,Cos2000b},
fourth \cite{BureauMII,Cos2000a},
or higher order \cite{ExtonRM} ODEs,
whose general solution is explicitly given because they have the PP.
Accordingly,
if one has an ODE in such an already studied class
(e.g.~second order second degree binomial-type ODEs \cite{CosScou}
$u''^2=F(u',u,x)$ with $F$ rational in $u'$ and $u$, analytic in $x$),
and which is suspected to have the PP
(for instance because one has been unable to detect any movable critical
singularity, see Section \ref{sectionPTest}),
then two cases are possible:
either there exists a transformation (\ref{eqHomographicGroup})
mapping it to a listed equation,
in which case the ODE has the PP and is explicitly integrated,
or such a transformation does not exist,
and the ODE has not the PP.

\section{Integrability and Painlev\'e property for
         partial differential equations}

Defining the PP for PDEs is not easy, 
but this must be done for future use in sections
\ref{sectionPTest} (the Painlev\'e test)
and 
\ref{sectionSingularity-based_methods} (proving the PP).
Such a definition must involve a \textit{global, constructive} property,
which excludes the concept of general solution. 
Indeed, this is only in nongeneric cases like the Liouville equation
that the general solution of a PDE can be built explicitly.
This is where the B\"acklund transformation comes in.
Let us first recall the definition of this powerful tool
(for simplicity, but this is not a restriction,
we give the basic definitions for a PDE defined as a single scalar equation
for one dependent variable $u$ and two independent variables $(x,t)$).

\begin{definition}
(Refs.~\cite[vol.~III chap.~XII]{DarbouxSurfaces}, \cite{MatveevSalle})
A \textbf{B\"acklund transformation}
(BT)
between two given PDEs
\index{B\"acklund transformation}
\begin{equation}
 E_1(u,x,t)=0,\ E_2(U,X,T)=0
\end{equation}
is a pair of relations
\begin{equation}
 F_j(u,x,t,U,X,T)=0,\ j=1,2
\label{eqBT}
\end{equation}
with some transformation between $(x,t)$ and $(X,T)$,
in which $F_j$ depends on the derivatives of $u(x,t)$ and $U(X,T)$,
such that the elimination of $u$ (resp.~$U$) between $(F_1,F_2)$
implies
$E_2(U,X,T)=0$ (resp.~$E_1(u,x,t)=0$).
In case the two PDEs are the same,
the BT is also called the \textbf{auto-BT}.
\index{B\"acklund transformation!auto--}
\end{definition}

Under a reduction PDE$\to$ODE,
the BT reduces to a birational transformation (also initials BT!),
which is \textit{not} involved in the definition of the PP for ODEs.
Therefore one needs an intermediate (and quite important) definition
before defining the PP.

\begin{definition}
A PDE in $N$ independent variables is \textbf{integrable}
if at least one of the following properties holds.
\index{integrability}
\begin{enumerate}
\item
Its general solution can be obtained, and it is an explicit closed form 
expression, possibly presenting movable critical singularities.

\item
It is linearizable. 

\item
For $N>1$, it possesses an auto-BT which, if $N=2$,
depends on an arbitrary complex constant, the B\"acklund parameter.
\index{B\"acklund parameter}

\item
It possesses a BT to another integrable PDE.

\end{enumerate}
\label{definitionIntegrable}
\end{definition}

Examples of these various situations are, respectively:
the PDE $u_x u_t + u u_{xt}=0$ 
with general solution $u=\sqrt{f(x)+g(t)}$,
which presents movable critical singularities 
and can be transformed into the d'Alembert equation;
the Burgers PDE $u_t + u_{xx} + 2 u u_x=0$, 
linearizable into the heat equation $\psi_t + \psi_{xx} =0$;
the KdV PDE $u_t + u_{xxx} - 6 u u_x=0$,
which is integrable by the inverse spectral transform (IST);
the Liouville PDE $u_{xt} + e^u=0$,
which possesses a BT to the d'Alembert equation $\psi_{xt}=0$.

We now have enough elements 
to give a definition of the Painlev\'e property 
for PDEs which is indeed an extrapolation of the one for ODEs.

\begin{definition}
The \textit{Painlev\'e property} (PP) of a PDE is
its integrability (definition \ref{definitionIntegrable})
and the absence of movable critical singularities 
near any noncharacteristic manifold.
\index{Painlev\'e!property}
\index{Painlev\'e!property for PDEs}
\label{definitionPPPDE}
\end{definition}

One will retain that the Painlev\'e property is a more demanding property
than the mere integrability.

The PP for PDEs is invariant under the natural extension 
of the homographic group (\ref{eqHomographicGroup}),
and \textit{classifications} similar to those of ODEs have also been performed
for PDEs,
 \index{classification}
in particular second order first degree PDEs 
\cite{CosPDEhyper,CosPDEpara},
isolating only the already known PDEs 
(Burgers, Liouville, sine-Gordon, Tzitz\'eica, etc).
Classifications based on other criteria,
such as the existence of an infinite number of conservation laws
\cite{MSS1991},
isolate more PDEs,
which are likely all integrable in the sense of 
definition \ref{definitionIntegrable};
it would be interesting to check that,
under the group of transformations generated by 
B\"acklund transformations and hodograph transformations,
each of them is equivalent to a PDE with the PP.
\index{hodograph transformation}

If one performs a hodograph transformation
(typically an exchange of the dependent and independent variables $u$ and $x$)
on a PDE having the PP,
the transformed PDE possesses a weaker form of the PP in which,
for instance, all leading powers and Fuchs indices become rational numbers
instead of integers.
Details can be found e.g. in Ref.~\cite{CFA}.
For instance, the Harry-Dym, Camassa-Holm \cite{CH1993}
and DHH equations \cite{DHH} can all be mapped to a PDE with the PP
by some hodograph transformation.

If the above definition of the PP for PDEs 
is really an extrapolation of the one for ODEs, 
then, given a PDE with the PP,
every reduction to an ODE which preserves the differential order
(i.e. a \textit{noncharacteristic reduction})
\index{reduction!noncharacteristic}
yields an ODE which necessarily has the PP.
This proves the conjecture of Ablowitz, Ramani and Segur \cite{ARS1980},
\textit{provided} the definition \ref{definitionPPPDE} 
really extrapolates that for ODEs, 
and this is the difficult part of this question.

There are plenty of such examples of reductions,
for instance the self-dual Yang-Mills equations
admit reductions to all six $\Pn$ equations \cite{MW1993}.

Given a PDE, the process to prove its integrability is twofold.
One must first check whether it may be integrable,
for instance by applying the \textit{Painlev\'e test}, 
as will now be explained in section \ref{sectionPTest}.
Then, in case of a nonnegative answer,
one must build explicitly the elements which are required to establish 
integrability,
for instance by using methods described in section
\ref{sectionSingularity-based_methods}.

Although partially integrable and nonintegrable equations,
i.e.~the majority of physical equations,
admit no BT, they retain part of the properties of (fully) integrable PDEs,
and this is why the methods presented below apply to both cases as well.
One such example is included below for information, 
section \ref{sectionKS}.

\section{The Painlev\'e test for ODEs and PDEs}
\label{sectionPTest}

 \index{Painlev\'e!test}

The test is an algorithm providing a set of \textit{necessary conditions}
for the equation to possess the PP.
Its full, detailed version can be found in
Ref.~\cite[\S 6]{Cargese1996Conte},
and below is only given a short presentation of its main subset,
known as the 
\textit{method of pole-like expansions},
 \index{method of pole-like expansions}
due to Kowalevski and Gambier \cite{GambierThese}.
The generated necessary conditions are \textit{a priori} not sufficient,
this was proven by Picard (1894), who exhibited the example
of the ODE with the general solution $u=\wp(\lambda \Log(x-c_1)+c_2,g_2,g_3)$,
namely
\begin{eqnarray}
& &
u'' - \frac{{u'}^2}{4 u^3 - g_2 u - g_3}\left(6 u^2 - \frac{g_2}{2}\right)
- \frac{{u'}^2}{\lambda \sqrt{4 u^3 - g_2 u - g_3}}=0,
\label{eqPicardEllipticExample}
\end{eqnarray}
which has the PP iff $2 \pi i \lambda$ is a period of the elliptic function
$\wp$,
a transcendental condition on $(\lambda,g_2,g_3)$ impossible to obtain
in a finite number of algebraic steps such as the Painlev\'e test.
Therefore it is wrong to issue statements like
``The equation passes the test, therefore it has the PP''.
The only way to prove the PP is,
\begin{description}
\item{--}
for an ODE,
either to explicitly integrate with the known functions
(solutions of linear, elliptic, hyperelliptic (a generalization of elliptic) 
or Painlev\'e equations),
or,
if one believes to have found a new function,
to prove both the absence of movable critical singularities and the
irreducibility \cite{U1990} to the known functions,
\index{irreducibility}

\item{--}
for a PDE,
to explicitly build the integrability elements of the definition
\ref{definitionPPPDE}.

\end{description}

Let us return to the test itself.
It is sufficient to present the method of pole-like expansions 
for one $N$th order equation 
\begin{equation}
    E(x,t,u)=0,
\label{eqDEgeneral}
\end{equation}
in one dependent variable $u$
and two independent variables $x,t$.
Movable singularities lay on a codimension one manifold
\begin{eqnarray}
& &
\varphi(x,t) - \varphi_0=0,
\label{eqPDEManifold}
\end{eqnarray}
in which the \textit{singular manifold variable}
              \index{singular manifold!variable}
$\varphi$ is an arbitrary function of the independent variables
and $\varphi_0$ an arbitrary movable constant.
Basically \cite{WTC},
the test of Weiss, Tabor and Carnevale
consists in checking the existence of all possible
local representations, near $\varphi(x,t) - \varphi_0=0$,
of the general solution
(whatever be its definition, difficult for PDEs)
as a locally single valued expression, e.g. the Laurent series
\begin{eqnarray}
& &
u = \sum^{+\infty}_{j=0} u_j \chi^{j+p},\ -p \in {\mathcal N},\
E = \sum^{+\infty}_{j=0} E_j \chi^{j+q},\ -q \in {\mathcal N},
\label{eqLaurentSeries}
\end{eqnarray}
with coefficients $u_j,E_j$ independent of the expansion variable $\chi$.
The natural choice $\chi=\varphi - \varphi_0$ \cite{WTC}
generates lengthy expressions $u_j,E_j$.
Fortunately, there is much freedom in choosing $\chi$,
only required to vanish as $\varphi - \varphi_0$ and to have a homographic
dependence on $\varphi - \varphi_0$ so as not to alter the structure of
movable singularities, hence the result of the test.
The unique choice which minimizes the
expressions and puts no constraint on $\varphi$ is \cite{Conte1989}
\begin{equation}
 \chi
=\frac{\varphi-\varphi_0}{\varphi_x
- \displaystyle{\frac{\varphi_{xx}}{2 \varphi_x}}
(\varphi-\varphi_0)}
=\left[\frac{\varphi_x}{\varphi-\varphi_0}
- \frac{\varphi_{xx}}{2 \varphi_x}\right]^{-1},\
\varphi_x \not=0,
\label{eqchi}
\end{equation}
in which $x$ denotes an independent variable whose component of
$\grad \varphi$ does not vanish.
The expansion coefficients $u_j,E_j$ are then invariant under the
six-parameter group of homographic transformations
\index{homographic!group}
\begin{eqnarray}
& &
\varphi \mapsto \frac{a' \varphi + b'}{c' \varphi + d'},\
a'd'-b'c' \not=0,
\end{eqnarray}
in which $a',b',c',d'$ are arbitrary complex constants,
and these coefficients only depend on the following elementary differential
invariants and their derivatives:
\index{homographic!invariants}
the \textit{Schwarzian}
     \index{Schwarzian}
\begin{eqnarray}
S=\lbrace \varphi;x \rbrace
=\frac{\varphi_{xxx}}{\varphi_x}
 - \frac{3}{2} \left(\frac{\varphi_{xx}} {\varphi_x} \right)^2,
\label{eqS}
\end{eqnarray}
and one other invariant per independent variable $t,y,\dots$
\begin{eqnarray}
C=- \varphi_t / \varphi_x,\
K=- \varphi_y / \varphi_x,\ \dots
\label{eqC}
\end{eqnarray}
The two invariants $S,C$ are linked by the cross-derivative condition
\begin{eqnarray}
X & \equiv &
((\varphi_{xxx})_t - (\varphi_t)_{xxx})/ \varphi_x
= S_t + C_{xxx} + 2 C_x S + C S_x = 0,
\label{eqCrossXT}
\end{eqnarray}
identically satisfied in terms of $\varphi$.

For the practical computation of $(u_j,E_j)$ as functions of $(S,C)$ only,
\ie\ what is called the
\textit{invariant Painlev\'e analysis},
 \index{Painlev\'e!analysis}
the variable $\varphi$ disappears,
and the \textit{only} information required is the gradient
of the expansion variable $\chi$,
\begin{eqnarray}
& &
\chi_x= 1 + \frac{S}{2} \chi^2,\
\chi_t= - C + C_x \chi  - \frac{1}{2} (C S + C_{xx}) \chi^2.
\label{eqChiGradient}
\end{eqnarray}
with the constraint (\ref{eqCrossXT}) between $S$ and $C$.

Consider for instance the
Kolmogorov-Petrovskii-Piskunov (KPP) equation
\cite{KPP,NewellWhitehead}
\index{KPP equation}
\begin{eqnarray}
& & E(u) \equiv
b u_t - u_{xx} + \gamma u u_x + 2 d^{-2} (u-e_1)(u-e_2)(u-e_3)=0,\
\label{eqKPP}
\end{eqnarray}
with $(b,\gamma,d^2)$ real and $e_j$ real and distinct,
encountered in reaction-diffusion systems
(the convection term $u u_x$ \cite{Satsuma1987}
is quite important in physical applications to prey-predator models).

The \textit{first step}, to search for the families of movable singularities
$u \sim u_0 \chi^p, E \sim E_0 \chi^q, u_0 \not=0$,
results in the selection of the \textit{dominant terms} $\hat E(u)$
\begin{eqnarray}
& &
\hat E(u) \equiv - u_{xx} + \gamma u u_x + 2 d^{-2} u^3,
\end{eqnarray}
which provide two solutions $(p,u_0)$
\begin{eqnarray}
& &
p=-1, q=-3, -2 - \gamma u_0 + 2 d^{-2} u_0^2=0.
\label{eqKPPFamilies}
\end{eqnarray}
The necessary condition that all values of $p$ be integer
is satisfied.

The second step is, for every selected family,
to compute the linearized equation,
\begin{eqnarray}
(\hat E'(u)) w
& \equiv &
\lim_{\varepsilon \to 0}
\frac{\hat E(u+\varepsilon w)-\hat E(u)}{\varepsilon}
\nonumber
\\
&=&(- \partial_x^2 + \gamma u \partial_x + \gamma u_x + 6 d^{-2} u^2) w=0,
\label{eqLinearized}
\end{eqnarray}
then its Fuchs indices $i$ near $\chi=0$
as the roots of the \textit{indicial equation}
\begin{eqnarray}
P(i)
{\hskip-3.0truemm}
& = &
\lim_{\chi \to 0} \chi^{-i-q}
(- \partial_x^2 + \gamma u_0 \chi^p \partial_x
  + \gamma p u_0 \chi^{p-1}+ 6 d^{-2} u_0^2 \chi^{2p})
 \chi^{i+p}
\\
{\hskip-3.0truemm}
& = &
  -(i-1)(i-2) + \gamma u_0 (i-2) + 6 d^{-2} u_0^2
\\
{\hskip-3.0truemm}
& = &
 -(i+1)(i-4- \gamma u_0)=0,
\end{eqnarray}
and finally to enforce the necessary condition that,
for each family, these two indices be distinct integers \cite{FP1991,CFP1993}.
Considering each family separately would produce a countable number of
solutions, which is incorrect.
Considering the two families simultaneously,
the
\textit{diophantine condition} that
 \index{diophantine condition}
the two values $i_1,i_2$ of the Fuchs index $4+\gamma u_0$ be integer
has a finite number of solutions, namely \cite[App.~I]{BureauMI}
\begin{eqnarray}
   & & \gamma^2 d^2=0,\   (i_1,i_2)=(4,4),\         u_0=(-d,d),
\\ & & \gamma^2 d^2=2,\   (i_1,i_2)=(3,6),\  \gamma u_0=(-1,2),
\\ & & \gamma^2 d^2=-18,\ (i_1,i_2)=(-2,1),\ \gamma u_0=(-6,-3).
\end{eqnarray}
It would be wrong at this stage to discard negative integer indices.
Indeed, in linear ODEs such as (\ref{eqLinearized}),
the single valuedness required by the Painlev\'e test restricts the
Fuchs indices to integers, whatever be their sign.
Let us proceed with the first case only, $\gamma=0$ (the usual KPP equation).

The recurrence relation for the next coefficients $u_j$,
\begin{equation}
\forall j\ge 1:\
E_j \equiv P(j) u_j + Q_j(\{u_l\ \vert \ l<j \}) = 0
\label{eqMethodPole5}
\end{equation}
depends linearly on $u_j$
and nonlinearly on the previous coefficients $u_l$.

The \textit{third and last step} is then to require,
for any admissible family and any Fuchs index $i$,
that the \textit{no-logarithm condition}
          \index{no-logarithm condition}
\begin{eqnarray}
& &
\forall i \in {\mathcal Z},\
P(i)=0 :\
Q_i=0
\label{eqConditionQ}
\end{eqnarray}
holds true.
At index $i=4$, the two conditions,
one for each sign of $d$ \cite{Conte1988},
\begin{eqnarray}
Q_4 & \equiv &
 C
 [(b d C + s_1 - 3 e_1)(b d C + s_1 - 3 e_2) (b d C + s_1 - 3 e_3)
\nonumber
\\
& &
\phantom{xxx} - 3 b^2 d^3 (C_t + C C_x)]
=0,\
s_1=e_1+e_2+e_3,
\label{eqKPPQ4}
\end{eqnarray}
are not identically satisfied, so the PDE fails the test.
This ends the test.

If instead of the PDE (\ref{eqKPP}) one considers its reduction
$u(x,t)=U(\xi), \xi=x-ct$ to an ODE,
then $C=\hbox{ constant}=c$,
and the two conditions $Q_4=0$ select the seven values
$c=0$ and $c^2=(s_1-3 e_k)^2 (b d)^{-2},k=1,2,3$.
For all these values, the necessary conditions are then sufficient since the
general solution $U(\xi)$ is singlevalued
(equation numbered 8 in the list of Gambier \cite{GambierThese}
reproduced in \cite{Ince}).

It frequently happens that the Laurent series (\ref{eqLaurentSeries}) 
only represents a particular solution,
for instance because some Fuchs indices are negative integers,
e.g.~the fourth order ODE \cite[p.~79]{BureauMII} 
\begin{equation}
 u'''' + 3 u u'' - 4 u'^2 = 0
\label{eqBureauOrder4}
\end{equation}
which admits the family
\begin{eqnarray}
& &
p=-2, u_0=-60, \hbox{ Fuchs indices } (-3,-2,-1,20).
\end{eqnarray}
The series (\ref{eqLaurentSeries}) depends on two, not four,
arbitrary constants, so two are missing and may contain multivaluedness.
In such cases,
one must perform a perturbation in order to represent the general solution
and to test the missing part of the solution for multivaluedness.

This perturbation \cite{CFP1993} is close to the identity
(for brevity, we skip the $t$ variable)
\begin{equation}
\label{eqPerturbu}
 x \hbox{ unchanged},\
 \bfu= \sum_{n=0}^{+ \infty} \varepsilon^n \bfu^{(n)}:\
 \bfE= \sum_{n=0}^{+ \infty} \varepsilon^n \bfE^{(n)}=0,
\end{equation}
where, like for the $\alpha-$method of Painlev\'e, 
the small parameter $\varepsilon$ is not in the original equation.

Then, the single equation (\ref{eqDEgeneral}) is equivalent to the infinite 
sequence
\begin{eqnarray}
\label{eqNL0} 
         n  =  0\ 
\bfE^{(0)}
{\hskip -3.5 truemm} & \equiv & {\hskip -3.5 truemm}
\bfE (x,\bfu^{(0)}) = 0,
\\
 \forall n \ge 1\ 
\bfE^{(n)}
{\hskip -3.5 truemm} & \equiv & {\hskip -3.5 truemm}
\bfE'(x,\bfu^{(0)}) \bfu^{(n)} 
             + \bfR^{(n)}(x,\bfu^{(0)},\dots,\bfu^{(n-1)}) = 0,
\label{eqLinn}
\end{eqnarray}
with $\bfR^{(1)}$ identically zero.
{}From a basic theorem of Poincar\'e 
\cite[Theorem II, \S 5.3]{Cargese1996Conte},
necessary conditions for the PP are
\begin{description} 
\item[-]
the general solution $\bfu^{(0)}$ of (\ref{eqNL0}) has no
movable critical points,
\item[-]
the general solution $\bfu^{(1)}$ of (\ref{eqLinn}) has no
movable critical points,
\item[-]
for every $n\ge 2$ there exists a particular solution of (\ref{eqLinn}) 
without movable critical points.
\end{description} 

Order zero is just the original equation (\ref{eqDEgeneral}) 
for the unknown $\bfu^{(0)}$,
so one takes for $\bfu^{(0)}$
the already computed (particular) Laurent series (\ref{eqLaurentSeries}).

Order $n=1$ is identical to the linearized equation 
\begin{equation}
 \bfE^{(1)} \equiv \bfE'(x,\bfu^{(0)}) \bfu^{(1)} = 0,
\label{eqLin0}
\end{equation}
and one must check the existence of $N$ independent solutions $\bfu^{(1)}$
locally singlevalued near $\chi=0$,
where $N$ is the order of (\ref{eqDEgeneral}).

The two main implementations of this perturbation are the 
Fuchsian perturbative method \cite{CFP1993}
and the
nonFuchsian perturbative method \cite{MC1995}.
In the above example (\ref{eqBureauOrder4}),                    
both methods indeed detect multivaluedness,
at perturbation order $n=7$ for the first one,
and $n=1$ for the second one (details below).

\subsection{The Fuchsian perturbative method}
\label{sectionMethodPerturbativeFuchsian}

Adapted to the presence of negative integer indices
in addition to the ever present value $-1$,
this method \cite{FP1991,CFP1993}
generates additional no-log conditions (\ref{eqConditionQ}).
Denoting $\rho$ the lowest integer Fuchs index, $\rho \le -1$,
the Laurent series for $\bfu^{(1)}$
\begin{equation}
 \bfu^{(1)}=\sum_{j=\rho}^{+ \infty} \bfu^{(1)}_j \chi^{j+\bfp},
\label{eqMethodPerturbative7}
\end{equation}
represents a particular solution containing a number of arbitrary coefficients
equal to the number of Fuchs indices, counting their multiplicity.
If this number equals $N$, it represents the general solution of
(\ref{eqLin0}).
Two examples will illustrate the method
\cite[\S 5.7.3]{Cargese1996Conte}.

The equation
\begin{equation}
u''+4 u u' + 2 u^3=0
\label{eqOrder2WithLog}
\label{eqdoublefamily}
\end{equation}
possesses the single family
\begin{eqnarray}
& &
p=-1,\
E_0^{(0)}= u_0^{(0)} (u_0^{(0)}-1)^2=0,\
\hbox{indices } (-1,0),
\end{eqnarray}
with the puzzling fact that $u_0^{(0)}$ should be at the same time 
equal to $1$ according to the equation $E_0^{(0)}=0$,
and arbitrary according to the index $0$.
The necessity to perform a perturbation arises from the multiple root
of the equation for $u_0^{(0)}$,
responsible for the insufficient number of arbitrary parameters in
the zeroth order series $u^{(0)}$.
The application of the method provides
\begin{eqnarray}
u^{(0)}
& = & 
\chi^{-1} \hbox{ (the series terminates)},
\\
E'(x,u^{(0)}) 
& = &
\partial_x^2 + 4 \chi^{-1} \partial_x + 2 \chi^{-2},
\\
u^{(1)}
& = &
u_{0}^{(1)} \chi^{-1},\
u_{0}^{(1)} \hbox{ arbitrary},\
\\
E^{(2)}
& = &
E'(x,u^{(0)}) u^{(2)} + 6 u^{(0)} u^{(1)^2} + 4 u^{(1)} u^{(1)'}
\nonumber
\\
& = &
\chi^{-2} (\chi^2 u^{(2)})'' + 2 u_0^{(1)^2} \chi^{-3}
=0,
\\
u^{(2)}
& = &
- 2 u_{0}^{(1)^2} \chi^{-1} (\Log \chi - 1).
\end{eqnarray}
The movable logarithmic branch point is therefore detected in a systematic way
at order $n=2$ and index $i=0$.
This result was of course found long ago 
by the $\alpha$-method \cite[\S 13, p 221]{PaiBSMF}.

The equation (\ref{eqBureauOrder4}) possesses the two families
\begin{eqnarray}
& &
{\hskip -16.0 truemm}
p=-2, u_0^{(0)}=-60, \hbox{ ind. } (-3,-2,-1,20),
  {\hat E}=u'''' + 3 u u'' - 4 u'^2, 
\\
& &
{\hskip -16.0 truemm}
p=-3, u_0^{(0)} \hbox{ arbitrary}, \hbox{ indices } (-1,0),
  {\hat E}= 3 u u'' - 4 u'^2. 
\label{eqBureau4p3}
\end{eqnarray}
 
The second family has a Laurent series $(p:+ \infty)$ which happens to
terminate \cite{CFP1993}
\begin{equation}
 u^{(0)}=c (x-x_0)^{-3}-60(x-x_0)^{-2},\ (c,x_0) \hbox{ arbitrary}.
\label{eqBureau4PartSol}
\end{equation}
For this family, the Fuchsian perturbative method is then useless,
because the two arbitrary coefficients corresponding to the two Fuchs indices
are already present at zeroth order.
 
The first family provides, at zeroth order, only a two-parameter
expansion and,
when one checks the existence of the perturbed solution
\begin{equation}
 u=\sum_{n=0}^{+ \infty} \varepsilon^n
 \left[\sum_{j=0}^{+ \infty} u_j^{(n)} \chi^{j-2-3n}\right],
\end{equation}
one finds that coefficients
$u_{20}^{(0)}, u_{-3}^{(1)}, u_{-2}^{(1)}, u_{-1}^{(1)}$
can be chosen arbitrarily,
and, at order $n=7$, one finds two violations \cite{CFP1993}
\begin{equation}
   Q_{-1}^{(7)} \equiv u_{20}^{(0)}   u_{-3}^{(1)^7} = 0,
   Q_{20}^{(7)} \equiv u_{20}^{(0)^2} u_{-3}^{(1)^6} u_{-2}^{(1)} = 0,
\end{equation}
implying the existence of a movable logarithmic branch point.
 
\subsection{The nonFuchsian perturbative method}
\label{sectionMethodPerturbativeNonFuchsian}

Whenever the number of indices is less than the differential order of
the equation,
the Fuchsian perturbative method fails to build a
representation of the general solution,
thus possibly missing some no-log conditions.
The missing solutions of the linearized equation (\ref{eqLin0})
are then solutions of the nonFuchsian type near $\chi=0$.

In section \ref{sectionMethodPerturbativeFuchsian},
the fourth order equation (\ref{eqBureauOrder4}) has been shown to fail
the test after a computation practically untractable without a computer.
Let us prove the same result without computation at all \cite{MC1995}.
The linearized equation
\begin{equation}
E^{(1)} = E'(x,u^{(0)}) u^{(1)} \equiv 
 [              \partial_x^4 
  + 3 u^{(0)}   \partial_x^2
  - 8 u^{(0)}_x \partial_x
  + 3 u^{(0)}_{xx}] u^{(1)} = 0,
\label{eqBureauLin}
\end{equation}
is known \textit{globally} for the second family because 
the two-parameter solution (\ref{eqBureau4PartSol}) is closed form,
therefore one can test all the singular points $\chi$ of (\ref{eqBureauLin}).
These are $\chi=0$ (nonFuchsian) and $\chi=\infty$ (Fuchsian),
and the key to the method is the information obtainable from $\chi=\infty$.
Let us first lower by two units the order of the linearized equation
(\ref{eqBureauLin}),
by ``subtracting'' 
the two global single valued solutions 
$u^{(1)}=\partial_{x_0} u^{(0)}$ and $\partial_c u^{(0)}$,
i.e.~$u^{(1)}=\chi^{-4},\chi^{-3}$,
\begin{equation}
u^{(1)}=\chi^{-4} v:\
[\partial_x^2 -16 \chi^{-1} \partial_x +3 c \chi^{-3} - 60 \chi^{-2}] v'' = 0,
\end{equation}
Then the local study of $\chi=\infty$ is unnecessary,
since one recognizes the Bessel equation.
The two other solutions in global form are
\begin{eqnarray}
c \not=0:\
v_1''
& = &
\chi^{-3} {}_{0} F_{1} (24;-3c/\chi)
=
\chi^{17/2} J_{23}(\sqrt{12 c/\chi}),
\\
v_2''
& = &
\chi^{17/2} N_{23}(\sqrt{12 c/\chi}),
\end{eqnarray}
where the hypergeometric function ${}_{0} F_{1} (24;-3c/\chi)$
is single valued and possesses an isolated essential singularity at $\chi=0$,
while the fonction $N_{23}$ of Neumann is multivalued because of a
$\Log \chi$ term.

\section{Singularity-based methods towards integrability}
\label{sectionSingularity-based_methods}

In this section,
we review a variety of singularity-based methods
able to provide some global elements of integrability.
The \textit{singular manifold method} of Weiss \textit{et al.} \cite{WTC} is
the most important of them, but it is not the only one.

A prerequisite notion is 
the \textit{singular part operator} ${\mathcal D}$,
     \index{singular part operator}
\begin{equation}
\Log \varphi \mapsto {\mathcal D} \Log \varphi=u_T(0)-u_T(\infty),
\label{eqSingularPartOperator}
\end{equation}
in which the notation $u_T(\varphi_0)$,
which emphasizes the dependence on $\varphi_0$,
stands for the principal part of the Laurent series
(\ref{eqLaurentSeries}),
\begin{eqnarray}
& &
u_T(\varphi_0) = \sum^{-p}_{j=0} u_j \chi^{j+p}.
\label{eqLaurentSeriesTruncated}
\end{eqnarray}
In our KPP example (\ref{eqKPPFamilies}) with $\gamma=0$,
this operator is
${\mathcal D}=d \partial_x$.

\subsection{Linearizable equations} 
\label{sectionLinearization}

When a nonlinear equation can be linearized,
the singular part operator
\index{singular part operator}
defined in (\ref{eqSingularPartOperator})
directly defines the linearizing transformation.

For instance, the Kundu-Eckhaus PDE
for the complex field $U(x,t)$ \cite{Kundu1984,CE1987}
\index{Kundu-Eckhaus equation}
\begin{eqnarray}
{\hskip -3.0 truemm}
& &
 i U_t + \alpha U_{xx}
+(\frac{\beta^2}{\alpha}{\mod{U}}^4 +2 b e^{i \gamma}({\mod{U}}^2)_x)U=0,\
(\alpha, \beta, b, \gamma) \in {\mathcal R},\
\label{eqKunduEckhaus}
\end{eqnarray}
with $\alpha \beta b \cos \gamma \not=0$,
passes the test iff \cite{CC1987,CMGalli1993} $b^2=\beta^2$.
Under the parametric representation 
\begin{eqnarray}
& &
U=\sqrt{u_x} e^{i \theta},
\end{eqnarray}
the equivalent fourth order PDE for $u$ \cite{CMGalli1993}
\begin{eqnarray}
& &
 \frac{\alpha}{2} (u_{xxxx} u_x^2 + u_{xx}^3 - 2 u_x u_{xx} u_{xxx})
+ 2 \frac{\beta^2 - (b \sin \gamma)^2}{\alpha} u_x^4 u_{xx}
\nonumber
\\
& &
+ 2 (b \cos \gamma) u_x^3 u_{xxx}
+\frac{1}{2 \alpha} (u_{tt} u_x^2 + u_{xx} u_t^2 - 2 u_t u_x u_{xt})
=0
\end{eqnarray}
admits two families, namely in the case $b^2=\beta^2$,
\begin{eqnarray}
& &
u \sim \frac{1}{2 \beta \cos \gamma} \Log \psi,\ \hbox{indices } -1,0,1,2,
\\
& &
u \sim \frac{3}{2 \beta \cos \gamma} \Log \psi,\ \hbox{indices } -3,-1,0,2,
\end{eqnarray}
in which $(\Log \psi)_x$ is the $\chi$ of the invariant Painlev\'e analysis.
When the test is satisfied ($b^2=\beta^2$),
the linearizing transformation \cite{Kundu1984}
is provided by \cite{CMGalli1993}
the singular part operator of the first family,
which maps the nonlinear PDE
(\ref{eqKunduEckhaus})
to the linear Schr\"odinger equation for $V$
obtained by setting $b=\beta=0$ in (\ref{eqKunduEckhaus}),
\begin{eqnarray}
& &
\hbox{Kundu-Eckhaus}(U)
\Longleftrightarrow
\hbox{Schr\"odinger}(V),\
 i V_t + \alpha V_{xx} = 0.
\\
& &
U=\sqrt{u_x} e^{i \theta},\
V=\sqrt{\varphi_x} e^{i \theta},\
u = \frac{\Log \varphi}{2 \beta \cos \gamma}.
\end{eqnarray}

\subsection{Auto-B\"acklund transformation of a PDE:
            the singular manifold method}
\label{sectionBT}

Widely known as the
\textit{singular manifold method} or \textit{truncation method}
because it selects
the beginning of a Laurent series and discards (``truncates'')
the remaining infinite part,
\index{truncation!method}
\index{singular manifold!method}
this method was introduced by Weiss \textit{et al.}~\cite{WTC}
and later improved in many directions
\cite{MC1991,EstevezEtAl1993,Garagash1993}
\cite{MC1994,CMGalli1995,Pickering1996SMM,MC1998}.
Its most recent version can be found in 
the lecture notes of a CIME school \cite{CetraroConte,CetraroMusette},
to which we refer for further details.

The goal is to find the B\"acklund transformation or,
\index{B\"acklund transformation}
if a BT does not exist,
to generate some exact solutions.
Since the BT is itself the result of an elimination \cite{Chen1974}
between the \textit{Lax pair} and the \textit{\formerDT},
\index{Darboux!involution}
\index{Lax pair}
the task splits into the two simpler tasks of deriving these two elements.
Let us take one example.

\index{Korteweg-de Vries equation!modified}

The modified Korteweg-de Vries equation (mKdV)
\begin{eqnarray}
& &
\hbox{mKdV}(w) \equiv
b w_t + \left(w_{xx} -2 w^3 / \alpha^2 \right)_x=0,\
\label{eqmKdV}
\end{eqnarray}
is equivalently written in its potential form 
\begin{eqnarray}
{\hskip -4.0 truemm}
& &
\hbox{p-mKdV}(r) \equiv
b r_t + r_{xxx} - 2 r_x^3 / \alpha^2 + F(t) = 0,\
w=r_x,
\end{eqnarray}
a feature which will shorten the expressions to come. 
This last PDE admits the two opposite families
($\alpha$ is any square root of $\alpha^2$)
\begin{eqnarray}
& &
p=0^{-},\
q=-3,\
r \sim \alpha \Log \psi,\
\hbox{indices } (-1,0,4),\
{\mathcal D} = \alpha,
\end{eqnarray}
and the results to be found are:
\begin{description}
\item -- the \formerDT
\begin{eqnarray}
& &
r={\mathcal D} \Log Y + R,\
\label{eqformerDT}
\end{eqnarray}
a relation expressing the difference of two solutions $r$ and $R$ 
of p-mKdV as the logarithmic derivative ${\mathcal D} \Log Y$,
in which ${\mathcal D} = \alpha$ is the singular part operator of either
family,
and $Y$ is a Riccati pseudopotential equivalent to the Lax pair
(see next item),

\item -- the Lax pair, here written in its equivalent Riccati representation,
\begin{eqnarray}
& &
\frac{y_x}{y}
=
\lambda (\frac{1}{y} - y) - 2 \frac{W}{\alpha},\
\label{eqmKdVRiccatix}
\\
& &
b \frac{y_t}{y}
=
 \frac{1}{y} \left(
- 4 \lambda \frac{W}{\alpha}
+ (2 \frac{W^2}{\alpha^2}
+ 2 \frac{W_{x}}{\alpha}
- 4 \lambda^2) y
 \right)_x,
\label{eqmKdVRiccatit}
\end{eqnarray}
in which $W$ satisfies the mKdV equation (\ref{eqmKdV})
and $\lambda$ is the spectral parameter,

\item -- the BT, by some elimination between the above two items.

\end{description}

This program is achieved by defining the truncation \cite{Pickering1996SMM},
\begin{eqnarray}
& &
r=\alpha \Log Y + R,\
\end{eqnarray}
in which $r$ satisfies p-mKdV,
$R$ is a yet unconstrained field,
$Y$ is the most general homographic transform of $\chi$ 
which vanishes as $\chi$ vanishes,
\begin{eqnarray}
& &
Y^{-1}=B(\chi^{-1} + A),\
\label{eqTruncationTwoFamily2}
\end{eqnarray}
$A$ and $B$ are two adjustable fields,
and the gradient of $\chi$ is (\ref{eqChiGradient}).
The \LHS\ of the PDE is then
\begin{eqnarray}
& &
\hbox{p-mKdV}(r) \equiv
\sum_{j=0}^{6} E_j(S,C,A,B,R) Y^{j-3},\
\end{eqnarray}
and the system of \textit{determining equations} to be solved is
\begin{eqnarray}
& &
\forall j\
E_{j}(S,C,A,B,R) =0.
\label{eqTruncationTwoFamily4}
\end{eqnarray}

This choice of $Y$ (\ref{eqTruncationTwoFamily2}) is necessary to implement
the two opposite families feature of mKdV.
The general solution of the determining equations 
introduces
an arbitrary complex constant $\lambda$ 
and a new field $W$
\cite{Pickering1996SMM}
\begin{eqnarray}
W & = & (R- \alpha \Log B)_x,\
A = W / \alpha,\
\nonumber
\\
b C & = & 2 W_{x} / \alpha -2 W^2 / \alpha^2 +4 \lambda^2,\
\nonumber
\\
S & = &   2 W_{x} / \alpha -2 W^2 / \alpha^2 -2 \lambda^2,\
\label{eqmKdVSCAB}
\end{eqnarray}
and the equivalence of the cross-derivative condition $(Y_x)_t=(Y_t)_x$ 
to the mKdV equation (\ref{eqmKdV}) for $W$
proves that one has obtained a \formerDT\ and a Lax pair,
with the correspondence $y=B Y$.

The auto-BT of mKdV
\index{B\"acklund transformation!auto--}
is obtained by the elimination of $Y$,
i.e. by the substitution
\begin{eqnarray}
& &
\Log (B Y) = \alpha^{-1} \int (w-W) \D x
\end{eqnarray}
in the two equations (\ref{eqmKdVRiccatix})--(\ref{eqmKdVRiccatit})
for the gradient of $ y=B Y$.

The \textit{singular manifold equation},
defined \cite{WTC} as the constraint put on $\varphi$ 
for the truncation to exist,
is obtained by the elimination of $W$ between $S$ and $C$,
\index{singular manifold!equation}
\begin{eqnarray}
& &
b C - S - 6 \lambda^2 = 0,
\end{eqnarray}
and it is identical to that of the KdV equation.
\index{Korteweg-de Vries equation}

\textit{Remark}.
The fact that,
in the Laurent series (\ref{eqLaurentSeriesTruncated}),
$u_T(0)$ (the ``\LHS'') and $u_T(\infty)$ (the ``constant level coefficient'')
are both solutions of the same PDE is not sufficient to define a BT,
since any nonintegrable 
PDE also enjoys this feature.
It is necessary to exhibit both the \formerDT\ and a good Lax pair.

Most $1+1$-dimensional PDEs with the PP have been successfully processed
by the singular manifold method,
including the not so easy 
Kaup-Kupershmidt \cite{MC1998} 
and Tzitz\'eica \cite{CMG1999} 
equations.

The extension to $2+1$-dimensional PDEs with the PP 
has also been investigated (Ref.~\cite{E2001} and references therein).

\subsection{Singlevalued solutions of the Bianchi IX cosmological model}
\label{sectionBianchiIX}

Sometimes, the no-log conditions generated by the test
provide some global information, which can then be used to integrate.

The Bianchi IX cosmological model 
is a 6-dim system of three second order ODEs
\begin{equation}
(\Log A)'' = A^2 - (B-C)^2
\hbox{ and cyclically},\
'=\D / \D \tau,
\label{eqBianchi1}
\end{equation}
or equivalently
\begin{equation}
(\Log \omega_1)''
= \omega_2^2 + \omega_3^2 - \omega_2^2 \omega_3^2 / \omega_1^2,\
A= \omega_2 \omega_3 / \omega_1,\
\omega_1^2=B C
\hbox{ and cyclically}.
\end{equation}

One of the families \cite{CGR1993,LMC1994}
\begin{eqnarray}
& &
A= \chi^{-1} + a_2 \chi + O(\chi^3),\ \chi=\tau-\tau_2,
\nonumber
\\
& &
B= \chi^{-1} + b_2 \chi + O(\chi^3),
\label{eqLaurentFamilyTwoOrder0}
\\
& &
C= \chi^{-1} + c_2 \chi + O(\chi^3),
\nonumber
\end{eqnarray}
has the Fuchs indices $-1,-1,-1,2,2,2$,
and the Gambier test detects no logarithms at the triple index $2$.
The Fuchsian perturbative method 
\begin{eqnarray}
& &
A=\chi^{-1} \sum_{n=0}^N \varepsilon^n
\sum_{j=-n}^{2+N-n} a_{j}^{(n)} \chi^{j},\
\chi=\tau-\tau_2,\
\hbox{ and cyclically},
\label{eqBianchi4}
\end{eqnarray}
then detects movable logarithms at $(n,j)=(3,-1)$ and $(5,-1)$ \cite{LMC1994},
and the enforcement of these no-log conditions generates the three solutions~:
\begin{eqnarray}
& &
(b_{ 2}^{(0)}=c_{ 2}^{(0)} \hbox{ and } b_{-1}^{(1)}=c_{-1}^{(1)})
\hbox{ or cyclically},
\label{eqSelectTaub}
\\
& &
a_{2}^{(0)}=b_{2}^{(0)}=c_{2}^{(0)}=0,\
\label{eqSelectHalphen}
\\
& &
a_{-1}^{(1)}=b_{-1}^{(1)}=c_{-1}^{(1)}.
\label{eqSelectEuler}
\end{eqnarray}
These are constraints which reduce the number of arbitrary coefficients
to, respectively, four, three and four,
thus defining particular solutions which may have no movable critical points.

The first constraint (\ref{eqSelectTaub})
implies the equality of two of the components $(A,B,C)$,
and thus defines the 4-dim subsystem $B=C$ \cite{Taub},
whose general solution is single valued,
\begin{eqnarray}
& &
A={k_1 \over \sinh k_1 (\tau-\tau_1)},\
B=C={k_2^2 \sinh k_1 (\tau-\tau_1) \over k_1 \sinh^2 k_2 (\tau-\tau_2)}.
\label{eqTaubNoMetric}
\end{eqnarray}

The second constraint (\ref{eqSelectHalphen}) 
amounts to suppress the triple Fuchs index $2$,
thus defining a 3-dim subsystem with a triple Fuchs index $-1$.
One can indeed check that the perturbed Laurent series (\ref{eqBianchi4})
is identical to that of the 
\textit{Darboux-Halphen system} \cite{Darboux1878,Halphen1881}
 \index{Darboux-Halphen system}
\begin{equation}
\omega_1' 
= \omega_2 \omega_3 - \omega_1 \omega_2 - \omega_1 \omega_3,\ 
\hbox{ and cyclically},
\label{eqDarboux}
\end{equation}
whose general solution is single valued.

The third and last constraint (\ref{eqSelectEuler})
amounts to suppress two of the three Fuchs indices $-1$,
thus defining a 4-dim subsystem whose explicit writing is yet unknown.
With the additional constraint
\begin{eqnarray}
& &
a_{2}^{(0)}+b_{2}^{(0)}+c_{2}^{(0)}= 0,
\label{eqSelectEulerRab}
\end{eqnarray}
the Laurent series (\ref{eqLaurentFamilyTwoOrder0}) is identical to that of
the 3-dim \textit{Euler system} (1750) \cite{BGPP},
describing the motion of a rigid body around its center of mass
\begin{equation}
\omega_1' = \omega_2 \omega_3,
\hbox{ and cyclically},
\label{eqEuler}
\end{equation}
whose general solution is elliptic \cite{BGPP}
\begin{eqnarray}
& &
{\hskip -5.0 truemm}
\omega_j= (\Log (\wp(\tau - \tau_0,g_2,g_3)-e_j))',\ j=1,2,3,\
(\tau_0,g_2,g_3) \hbox{ arbitrary},
\\
& &
{\hskip -5.0 truemm}
{\wp'}^2=4 (\wp-e_1)(\wp-e_2)(\wp-e_3)=4 \wp^3 - g_2 \wp - g_3.
\end{eqnarray}
The 4-dim subsystem (the one without (\ref{eqSelectEulerRab}))
defines an extrapolation to four parameters of this elliptic solution,
quite probably single valued, whose closed form is still unknown.

One thus retrieves by the analysis all the results of the geometric 
assumption of self-duality \cite{GibbonsPope1979},
even slightly more.

\subsection{Polynomial first integrals of a dynamical system}
\label{sectionLorenzFI}

A first integral of an ODE is by definition a function of $x,u(x),u'(x),\dots$
which takes a constant value at any $x$,
including the movable singularities of $u$.
Consider for instance the
\textit{Lorenz model}
 \index{Lorenz model}
\begin{equation}
\frac{\D x}{\D t} = \sigma (y-x),\
\frac{\D y}{\D t} = r x - y - x z,\
\frac{\D z}{\D t} = x y - b z (x-y).
\label{eqLorenz}
\end{equation}

First integrals in the class $P(x,y,z) e^{\lambda t}$,
with $P$ polynomial and $\lambda$ constant,
should not be searched for with the assumption $P$ the most general polynomial
in three variables.
Indeed, $P$ must have no movable singularities.
The movable singularities of $(x,y,z)$ are
\begin{eqnarray}
& &
x \sim  2 i             \chi^{-1},\
y \sim -2 i \sigma^{-1} \chi^{-2},\
z \sim -2   \sigma      \chi^{-2},\
\hbox{indices } (-1,2,4),
\end{eqnarray}
therefore the generating function of admissible polynomials $P$
is built from the singularity degrees of $(x,y,z)$ \cite{LevineTabor}
\begin{eqnarray}
& &
{\hskip -10.0 truemm}
\frac{1}{(1 - \alpha x) (1 - \alpha^2 y) (1 - \alpha^2 z)}= 
1 + \alpha x + \alpha^2 \left(x^2+y+z \right)
  + \alpha^3 \left(x^3+xy+xz \right)
\nonumber
\\
& &
{\hskip -10.0 truemm}
\phantom{
\frac{1}{(1 - \alpha x) (1 - \alpha^2 y) (1 - \alpha^2 z)}
}
  + \alpha^4 \left(x^4+x^2 y+ x^2 z+ y z+ z^2+ y^2 \right)
+ \dots
\end{eqnarray}
defining the basis, ordered by singularity degrees,
\begin{eqnarray}
& &
(1),\
(x),\
(x^2,y,z),\
(x^3,x y, x z),\
(x^4,x^2 y, x^2 z, y z, z^2, y^2),\
\dots
\end{eqnarray}
The candidate of lowest degree is a linear combination of $(x^2,y,z)$,
which indeed provides a first integral \cite{Segur}
\begin{eqnarray}
& &
K_1 = (x^2 - 2 \sigma z) e^{2 \sigma t},\ b=2 \sigma.
\end{eqnarray}
Six polynomial first integrals are known \cite{Kus} with a singularity degree
at most equal to four,
and these are the only ones \cite{LlibreZhang}.

\subsection{Solitary waves from truncations}
\label{sectionKS}

If the PDE is nonintegrable or if one only wants to find particular solutions,
the singular manifold method of Section \ref{sectionBT} still applies,
it simply produces less results.
For autonomous partially integrable PDEs,
the typical output is a set of constant values for the unknowns
$S,C,A,B,R$ in the determining equations (\ref{eqTruncationTwoFamily4}).
In such a case,
quite generic for nonintegrable equations, 
the integration of the Riccati system (\ref{eqChiGradient})
yields the value
\begin{eqnarray}
& &
\chi^{-1}=\frac{k}{2} \tanh \frac{k}{2} (\xi-\xi_0),\
\xi=x-ct,\ k^2=-2 S,\ c=C,
\end{eqnarray}
the singular part operator $\mathcal{D}$ has constant coefficients,
therefore the solutions $r$ in (\ref{eqformerDT}) 
are
\textit{solitary waves} $r=f(\xi)$,
 \index{solitary waves}
in which $f$ is a polynomial
in $\sech k \xi$ and $\tanh k \xi$.
This follows immediately from 
the two elementary identities \cite{CM1993}
\begin{equation}
   \tanh z - \frac{1}{\tanh z}= -2 i \sech\left[2 z + i \frac{\pi}{2}\right],\
   \tanh z + \frac{1}{\tanh z}=  2   \tanh\left[2 z + i \frac{\pi}{2}\right].
\label{eqIdentitiestanhsech}
\end{equation}

In the simpler case of a one-family PDE, 
the (degenerate) \formerDT\ is
\begin{eqnarray}
& &
u={\mathcal D} \Log \psi + U,\
\partial_x \Log \psi=\chi^{-1},
\end{eqnarray}
and the above class of solitary waves $r=f(\xi)$ reduces to the class of
polynomials in $\tanh (k/2) \xi$.
In the example of the chaotic Kuramoto-Sivashinsky (KS) equation
\index{Kuramoto-Sivashinsky equation}
\begin{eqnarray}
& &
u_t + u u_x + \mu u_{xx} + b u_{xxx} + \nu u_{xxxx}=0,\ \nu \not=0,
\label{eqKS}
\end{eqnarray}
one finds \cite{KudryashovKSFourb} 
\begin{eqnarray}
& &
\mathcal{D}=60 \nu \partial_x^3 + 15 b \partial_x^2
           + \frac{15(16 \mu \nu - b^2)}{76 \nu} \partial_x,
\label{eqKSD}
\\
& &
u=\mathcal{D} \Log \cosh \frac{k}{2} (\xi-\xi_0) + c,\ 
(c,\xi_0) \hbox{ arbitrary},
\label{eqKSresult}
\end{eqnarray}
in which $b^2/(\mu \nu)$ only takes the values $0,144/47,256/73,16$,
and $k$ is not arbitrary.
In the quite simple writing (\ref{eqKSresult}),
much more elegant than a third degree polynomial in $\tanh$,
the only nonlinear item is the logarithm,
$\mathcal{D}$ being linear and $\cosh$ solution of a linear system.
This displays the enormous advantage to take into account the
singularity structure when searching for such solitary waves.

See e.g. Ref.~\cite{CM2000b} for a recent application to coupled
Ginzburg-Landau equations.

\subsection{First degree birational transformations of Painlev\'e equations}

At first glance,
it seems that the truncation procedure described in section \ref{sectionBT}
should be even easier when the PDE reduces to an ODE.
This is not the case, because, in addition to the Riccati variable
$\chi$ or $Y$ of the truncation,
there exists a second natural Riccati variable
and therefore a homographic dependence between the two Riccati variables,
which must be taken into account under penalty of failure of
the truncation.

Indeed,
any $N$th order, first degree ODE with the Painlev\'e property
is necessarily \cite[pp.~396--409]{PaiLecons} a
Riccati equation for $U^{(N-1)}$, with coefficients depending on
$x$ and the lower derivatives of $U$, e.g. in the case of $\PVI$,
\index{Painlev\'e!equation $\PVI$}
\begin{eqnarray}
& & U''=A_2(U,x) U'^2 + A_1(U,x) U' + A_0(U,x,\Alpha,\Beta,\Gamma,\Delta).
\label{eqRiccatiUprime}
\end{eqnarray}
Then the Riccati variable of the truncation (denote it $Z$)
is linked to $U'$ by some homography,
\begin{eqnarray}
& &
 (U' + g_2) (Z^{-1} - g_1) - g_0=0,\ g_0 \not=0,
\label{eqHomographyRUprime}
\end{eqnarray}
in which $g_0,g_1,g_2$ are functions of $(U,x)$ to be found.
Implementing this dependence in the truncation \cite{CM2001b}
provides a unique solution for $\PVI$,
which is 
the unique first degree birational transformation, 
first found by Okamoto \cite{Okamoto1986Pn},
\index{Painlev\'e!equation $\PVI$}
\begin{eqnarray}
\frac{N}{u-U}
& = &
  \frac{x (x-1) U'}{U (U-1)(U-x)}
 +\frac{\Theta_0}{U}+\frac{\Theta_1}{U-1}+\frac{\Theta_x-1}{U-x}
\label{eqTP6uvecTUnsigned}
\\
& = &
  \frac{x(x-1)u'}{u(u-1)(u-x)}
 +\frac{\theta_0}{u}+\frac{\theta_1}{u-1}+\frac{\theta_x-1}{u-x},
\label{eqTP6uvectUnsigned}
\\
\forall j=\infty,0,1,x\ & : & \
(\theta_j^2+ \Theta_j^2 - (N/2)^2)^2 - (2 \theta_j \Theta_j)^2=0,
\label{eqP6AlgebraicBira}
\\
N
& = &
\sum (\theta_k^2 - \Theta_k^2),
\label{eqP6N}
\end{eqnarray}
with the classical definition for the monodromy exponents,
\begin{eqnarray}
& &
\theta_\infty^2= 2 \alpha,\
\theta_0^2     =-2 \beta,\
\theta_1^2     = 2 \gamma,\
\theta_x^2     =1 - 2 \delta,
\\
& &
\Theta_\infty^2= 2 \Alpha,\
\Theta_0^2     =-2 \Beta,\
\Theta_1^2     = 2 \Gamma,\
\Theta_x^2     =1 - 2 \Delta.
\end{eqnarray}
The equivalent affine representation of
(\ref{eqP6AlgebraicBira})--(\ref{eqP6N}) is
\begin{eqnarray}
\theta_j &=& \Theta_j - \frac{1}{2} \left(\sum \Theta_k\right) + \frac{1}{2}
\ccomma\
\Theta_j = \theta_j - \frac{1}{2} \left(\sum \theta_k\right) + \frac{1}{2}
\ccomma
\label{eqT9}
\\
N
& = &
  1 - \sum \Theta_k
=
 -1 + \sum \theta_k
=
  2 (\theta_j - \Theta_j),\ j=\infty,0,1,x,
\label{eqNvecT}
\end{eqnarray}
in which $j,k=\infty,0,1,x$.

The well known confluence from $\PVI$ down to $\PII$
then allows us to recover \cite{CM2001c} all the first degree
birational transformations of the five Painlev\'e equations
($\PI$ admits no such transformation because it depends on no parameter),
thus providing a unified picture of these transformations.

\section{Liouville integrability and Painlev\'e integrability}

A Hamiltonian system with $N$ degrees of freedom is said
\textit{Liouville--integrable}
 \index{Liouville--integrability}
if it possesses $N$ functionally independent invariants in involution.
In general, there is no correlation between 
Liouville--integrability and the Painlev\'e property,
as seen on the two examples with $N=1$
\begin{eqnarray}
& &
H(q,p,t)=\frac{p^2}{2} - 2 q^3 - t q,\
\\
& &
H(q,p,t)=\frac{p^2}{2} + q^5,
\end{eqnarray}
in which 
the first system is Painlev\'e-integrable and not Liouville--integrable,
and \textit{vice versa} for the second system.
However,
given a Liouville--integrable Hamiltonian system,
which in addition passes the Painlev\'e test,
one \textit{must} try to prove its Painlev\'e integrability
by explicitly integrating.

Such an example is the cubic H\'enon-Heiles system
\index{H\'enon-Heiles system}
\begin{eqnarray}
H
& \equiv & 
 \frac{1}{2} (p_1^2 + p_2^2 + c_1 q_1^2 + c_2 q_2^2)
    + \alpha q_1 q_2^2 - \frac{1}{3} \beta q_1^3 + \frac{1}{2} c_3 q_2^{-2},
\alpha \not=0,
\\
& & q_1'' + c_1 q_1 - \beta q_1^2 + \alpha q_2^2 = 0,
\label{eqHH1}
\\
& & q_2'' + c_2 q_2 + 2 \alpha q_1 q_2 - c_3 q_2^{-3} 
=0,
\label{eqHH2}
\end{eqnarray}
which passes the Painlev\'e test in three cases only,
\begin{eqnarray}
\hbox{(SK)} : & & 
\beta/ \alpha=-1,
c_1=c_2,\
\\
\hbox{(K5)} : & & 
\beta/ \alpha=-6,\
\\
\hbox{(KK)} : & & 
\beta/ \alpha=-16,\
c_1=16 c_2.
\end{eqnarray}
In these three cases, the general solution $q_1$ (hence $q_2^2$)
is indeed singlevalued and expressed with genus two hyperelliptic
functions.
This was proven by Drach in 1919 for the second case,
associated to KdV${}_5$,
and only recently 
\cite{VMC2002a}
in the two other cases.
This proof completes the result of Ref.~\cite{RGC},
who found 
the \textit{separating variables} (a global object)
by just considering the Laurent series (a local object),
following a powerful method due to van Moerbeke and Vanhaecke
\cite{VanhaeckeLNM}.
 \index{separating variables}

\section{Discretization and discrete Painlev\'e equations}

This quite important subject (the integrability of difference equations)
is reviewed elsewhere in this volume
\cite{pinsaDiscrete},
so we will just write a few lines about it, for completeness.

Let us consider the difference equations or $q$-difference equations
(we skip for shortness the elliptic stepsize \cite{Sakai1999}),
\begin{eqnarray}
& &
\forall x\ \forall \pas\ :\
E(x,\pas,\{u(x+k \pas),\ k-k_0=0,\dots,N\})=0,
\label{eqDiscretexpas}
\\
& &
\forall x\ \forall \pasq\ :\
E(x,\pasq,\{u(x \pasq^k),\ k-k_0=0,\dots,N\})=0,
\label{eqDiscretexpasq}
\end{eqnarray}
algebraic in the values of the field variable,
with coefficients analytic in $x$ and the stepsize $\pas$ or $\pasq$.
As compared to the continuous case,
the main missing item is an undisputed definition for the 
\textit{discrete Painlev\'e property}.
 \index{Painlev\'e!property!discrete}
The currently proposed definitions are

\begin{enumerate}
\item
\cite{CM1996}
There exists a
neighborhood of $\pas=0$ (resp.~$\pasq=1$)
at every point of which
the general solution $x \to u(x,\pas)$ (resp.~$x \to u(x,\pasq)$)
has no movable critical singularities.

\item
\cite{AHH}
The Nevanlinna order of growth of the solutions at infinity is finite.

\end{enumerate}
but none is satisfactory.
Indeed,
the first one says nothing about discrete equations without continuum limit,
and
the second one excludes the continuous $\PVI$ equation.

Despite the lack of consensus on this definition,
a
\textit{discrete Painlev\'e test}
 \index{Painlev\'e!test!discrete}
has been developed to generate necessary conditions
for the above properties.
Of exceptional importance at this point is the
\textit{singularity confinement method} \cite{GRP1991},
 \index{singularity confinement method}
which tests with great efficiency a property not yet rigorously defined,
but which for sure will be an important part of the good definition of the
discrete Painlev\'e property.
The approach developed by Ruijsenaars \cite{Ruijsenaars}
for linear discrete equations,
namely to require as much analyticity as possible,
should be interesting to transpose to nonlinear discrete equations.

Just for consistency,
an interesting development would be to display a discrete version of
(\ref{eqPicardEllipticExample})
escaping all the methods of the discrete test.

Let us say a word on the discrete analogue of the Painlev\'e and Gambier
classification.
These second order first degree continuous equations all have a precise form
($u''$ is a second degree polynomial in $u'$,
the coefficient of ${u'}^2$ is the sum of at most four simple poles in $u$,
etc),
directly inherited from the property of the elliptic equations isolated by
Briot and Bouquet.
In the discrete counterpart,
the main feature is the existence 
of an addition formula for the elliptic function $\wp$ of Weierstrass.
As remarked earlier by Baxter and Potts (see references in \cite{CM1998}),
this formula defines an \textit{exact discretization} of 
(\ref{eqWeierstrassOrder1}).
Then,
all the autonomous discrete second order first degree equations with
the (undefined!) discrete PP 
have a precise form resulting from the most general discrete differentiation
of the addition formula,
and the nonautonomous ones simply inherit variable coefficients as in the
continuous case.
Of course, the second order higher degree (mostly multi-component)
equations are much richer,
see details in the review \cite{Cargese1996GNR}.

Another open question concerns the continuum limit of the 
\textit{contiguity relation} of the ODEs which admit such a relation.
 \index{contiguity relation}
The contiguity relation of the (linear) hypergeometric equation
has a continuum limit which is not the hypergeometric equation, 
but a confluent one.
On the contrary, 
the contiguity relation of the (linearizable) Ermakov equation
has a continuum limit which is again an Ermakov equation \cite{Hone1999}.
One could argue that the latter depends on a function,
and the former only on a finite number of constants.
Nevertheless,
this could leave the hope to upgrade from $\PV$ to $\PVI$ the highest
continuum limit for the 
contiguity relation of $\PVI$ \cite{Okamoto1987I,NRGO,CM2001c}.

\section{Conclusion}

The allowed space forced us to skip quite interesting developments,
such as the relation with differential geometry \cite{BE2000},
or the way to obtain the nonlinear superposition formula
from singularities \cite{MV2000b},
or the weak Painlev\'e property 
\cite[\Lecons\ 5--10,13,19]{PaiLecons} \cite{RDG,GDR1984b}.

For applications to nonintegrable equations,
not covered in this Special issue,
the reader can refer to tutorial presentations such that
\cite{CetraroConte,Cargese1996Musette}.

\section*{Acknowledgments}

The financial support of the Tournesol grant T99/040,
the IUAP Contract No.~P4/08 funded by the Belgian government,
and the CEA, is gratefully acknowledged.

\vfill \eject

\printindex


\vfill \eject
\end{document}